\documentclass{PoS}

\newcommand{\tr}{\mbox{tr}}

\newcommand{\diag}{\mbox{diag}}
\newcommand{\Id}{\mbox{Id}}
\newcommand{\Od}{{\cal O}}

\title{Pion masses at finite temperature}

\ShortTitle{}

\author{\speaker{R. Torres Andres}\\
        Departamento de F\'{\i}sica
Te\'orica II, Universidad Complutense, 28040 Madrid, Spain\\
        E-mail: \email{rtandres@fis.ucm.es}}

\author{A. Gomez Nicola\\Departamento de F\'{\i}sica
Te\'orica II, Universidad Complutense, 28040 Madrid, Spain\\
       E-mail: \email{gomez@fis.ucm.es}}
\abstract{We present preliminary results on a study about the thermal variation of the charged and neutral pion masses to one loop, analyzing their electromagnetic difference, in the context of Chiral Perturbation Theory with two flavours, as well as using a light resonance model. We find that the pion mass difference increases for, at least, low and intermediate temperatures, unlike the chiral limit decreasing result. The axial-vector mixing arising from chiral restoration smooths the Debye-screening temperature increase. Taking into account further corrections due to axial and vector resonances, dominated by a$_1$ and  $\rho$ particles respectively, does not change significantly the ChPT prediction.}

\FullConference{Xth Quark Confinement and the Hadron Spectrum,\\
		October 8-12, 2012\\
		TUM Campus Garching, Munich, Germany}

\begin{document}

\section{Introduction}

 SU(2)-Chiral perturbation theory (SU(2)-ChPT) is one of the most powerful techniques to describe the low energy sector of the strong interaction. It is an effective field theory based on the spontaneous breaking of the chiral group, $SU_V(2)\times SU_A(2)\rightarrow SU_V(2)$, and provides a systematic and model-independent scheme for calculating low energy observables at zero  \cite{we79,Gasser:1983yg,Gasser:1984gg} and also at finite temperature \cite{Gasser:1986vb,Gerber:1988tt}, which is very important to compare with the results provided by lattice calculations \cite{Borsanyi:2012cr,Bazavov:2011nk} below the pseudocritical temperature (180-200 MeV).

  The terms of the effective Chiral Lagrangian are constructed as an expansion in even powers of a meson energy scale (external momentum, meson mass) which is to be compared finally with the chiral scale, $\Lambda\sim 1$ GeV, that separates in an approximate way the region of high energy processes from the low energy ones.

  In order to implement electromagnetic (EM) effects, the vector subgroup $SU_V(2)$ must be explicitly broken by adding a photon field and a quark charge matrix via the external source method, giving rise to new terms in the Chiral Lagrangian \cite{Urech:1994hd,Knecht:1997jw,Meissner:1997fa} and introducing the need of a change in the chiral counting scheme in order to accommodate the electric charge, $e$, which is formally achieved considering $e^2=\mathcal{O}(p^2/F^2)$, being $F$ the pion decay constant in the chiral limit.

We present here preliminary results for the leading order corrections to the real part of the self-energy of neutral and charged pions propagating in a thermal bath. Our analysis could be useful when comparing neutral and charged pion distributions observed in Heavy Ion Collisions. Actually, recent experimental data are able to measure and compare charged and pion distributions very accurately \cite{Abelev:2009wx,ConesaBalbastre:2011zi}, showing that their difference is negligible within errors at very low $p_T$,  where our approach would be applicable. Besides, there are finite temperature lattice analysis  measuring pion screening masses \cite{Cheng:2010fe} which should not differ much from the pole masses analyzed here  below but not close to the transition.  EM interactions are customarily not included in lattice simulations and therefore an study of their effect in different thermal quantities is interesting in order to support or not such approximation. Finally, the analysis has also formal interest, since the pion mass difference is connected to the axial-vector spectral function difference at $T=0$ \cite{Das:1967it} and at $T\neq 0$ in the chiral limit \cite{Manuel:1998sy}. Since non-zero quark masses invalidate by definition the use of the soft pion limit leading to such connection, the analysis for non-zero masses and finite temperature is pertinent and the role of resonances can be established by the use of effective models. The detailed analysis, results and discussion will be the subject of a forthcoming work \cite{TorresNicola:2013}.

The expression for the pole of the pion propagator establishes the dispersion relation as customary, as  $p_0^2=\vec p^2+\hat M^2_\pi+\Sigma_\pi(p_0,\vec p)$
where $\vec p$ is the three-momentum of the external particle, $\hat M_\pi$ is the tree level mass of the pion, and $\Sigma_\pi$ is the $\Od(p^4)$ pion self-energy, which encodes all the interactions with the medium and will depend separately on $p_0$ and $\vec p$ as a result of the Lorentz Symmetry breaking due to the choice of the reference system associated with the thermal bath. If EM isospin breaking is considered, the dispersion relation is different for charged and neutral pions already at tree level. The real part of $\Sigma_\pi$ gives information about the dispersion relation and the pion velocity, whereas the imaginary part is related to the absorption in the medium. Since we are interested only in calculating masses, we will focus our study in the real part and will make use of the position of the pole of the in-medium pion propagator, i.e, our modified masses of charged and neutral pions will be the solutions of
\begin{equation}
M_{\pi^i}^2(\vec{p})-\hat M_{\pi^i}^2-Re\Sigma_{\pi^i}(p_0^2=\vec{p}^2+\hat M_{\pi^i}^2,\vec{p}) =0,\hspace{4mm} i=\{0,\pm\},
\end{equation}
where the self-energy is calculated perturbatively in ChPT and therefore is on-shell, with a residual thermal three-momentum dependence. We will define, as customary, the pion masses in the static limit $\vec{p}=\vec{0}$.

\section{Pion masses in SU(2) ChPT}

The effective chiral lagrangian up to fourth order in $p$ (a meson mass, momentum, temperature or derivative) including electromagnetic interactions proportional to $e^2$ is given schematically by ${\cal L}_{eff}={\cal L}_{p^2+e^2}+{\cal L}_{p^4+e^2p^2+e^4}$.  The second order lagrangian corresponds to the familiar non-linear sigma model plus the addition of the gauge coupling of mesons to the photon field via the covariant derivative, and an additional term proportional to a low-energy constant $C$ compatible with the $e\neq 0$ symmetries of the QCD lagrangian \cite{Urech:1994hd,Ecker:1988te}:
\begin{equation}
{\mathcal L}_{p^2+e^2}=\frac{F^2}{4} \tr\left[D_\mu U^\dagger D^\mu U+2B_0{\cal M}\left(U+U^\dagger\right)\right]+C\tr\left[QUQU^\dagger\right]
\label{L2},
\end{equation}
where $U(x)=\exp [i\Phi/F]\in SU(2)$, being $\Phi$ the pion matrix field; and the covariant
derivative is defined through $D_\mu=\partial_\mu+iA_\mu[Q,\cdot]$ with $A_\mu$ the photon field. ${\cal M}$ and $Q$ are the quark mass and charge matrices defined by ${\cal M}=m \Id$ and
$Q=(e/3)\diag(2,-1)$, and $C=(\hat M^2_{\pi^\pm}-\hat M^2_{\pi^0})F^2/2e^2$.
The fourth order lagrangian consists of all possible terms compatible with the QCD symmetries to that order, including the EM ones, and can be found for SU(2)-ChPT, for example, in  \cite{Knecht:1997jw}. We will not have to be worried by the set of EM and non EM low energy constants introduced by ${\cal L}_{p^4+e^2p^2+e^4}$, for we will be mainly interested in thermal contributions and the fourth order lagrangian is just used to absorb the $T=0$ divergences coming from the loops. The complete zero temperature one-loop values for the pion masses including EM corrections can be found for instance in \cite{Knecht:1997jw,TorresAndres:2011wd}.

\begin{figure}
\centering
\includegraphics[scale=0.08]{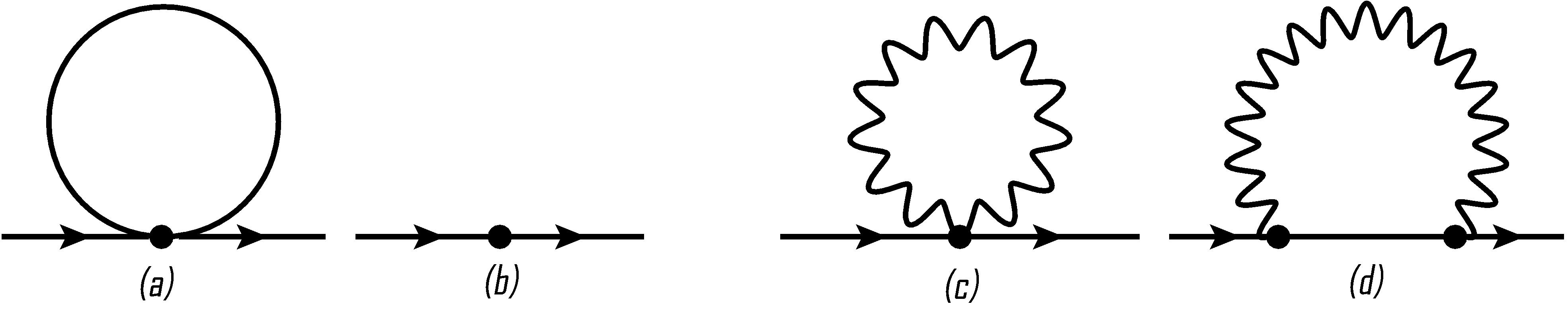}
\caption{1-PI diagrams contributing to the self-energy of a charged pion in SU(2)-ChPT to leading order. Diagrams for neutral pions are the same removing those in which photons are present.}
\label{fig:diag}
\end{figure}

The self-energy of a charged pion to leading order consists of all the 1-particle irreducible (1-PI) diagrams modifying the pion propagator at $\mathcal{O}(p^4)$ in SU(2)-ChPT , shown in Fig.\ref{fig:diag}. The neutral pion does not feel the electromagnetic field to this order, so the corrections to its mass do not involve photon fields (the rest of the diagrams are the same although the vertices are different).

The photon-tadpole diagram Fig.\ref{fig:diag}(c) is proportional to the photon mass and therefore vanishes at zero temperature, while pion tadpoles Fig.\ref{fig:diag}(a) and the photon-exchange diagram Fig.\ref{fig:diag}(d) are finite and chiral scale-independent once regularized and combined with diagram Fig.\ref{fig:diag}(b).

The pion tadpoles, charged or not, give rise to $g_1(M,T)$ thermal functions defined as
$
\!g_i(M^2_i,T)\!=\frac{1}{2\pi^2}\int_0^\infty dp \frac{p^2}{E_p} \frac{1}{e^{\beta E_p}-1},
$
with $E_p^2=p^2+M_i^2$ and $\beta=T^{-1}$. For the neutral pion mass we get
\begin{equation}
M^2_{\pi^0}=\hat M^2_{\pi^0}(T=0)\left[1+\frac{1}{F^2}\left(g_1(\hat M^2_{\pi^\pm},T)-\frac{1}{2}g_1(\hat M^2_{\pi^0},T)\right)\right].
\end{equation}

The above correction due to the EM tree masses correction represents numerically only a slight deviation from the $\hat M^2_{\pi^\pm}=\hat M^2_{\pi^0}$ limit, the neutral pion mass still increasing with temperature.

The charged pion mass receives contributions from two different sources: pion tadpoles and virtual photon diagrams, which also can be of two different types: a photon-tadpole contribution (Fig.\ref{fig:diag}(c)) which grows like $e^2T^2$, just in the same manner as a Debye or screening mass of the electric field in a thermal bath \cite{Kraemmer}; and a more complicated  structure of one-photon exchange (Fig.\ref{fig:diag}(d)) that depends on the external pion momentum. Putting together all the pieces we get
\begin{equation}
M^2_{\pi^\pm}(\vec{p})=\hat M^2_{\pi^\pm}-4Ze^2g_1(\hat M^2_{\pi^\pm},T)+\frac{1}{3}e^2T^2+Re\Sigma_{\gamma Ex}(\vec{p}),
\end{equation}
being $p$ the external pion three-momentum, $Z=C/F^4$  and $\Sigma_{\gamma Ex}(\vec p)$ the self energy contribution from the one-photon exchange on the mass shell \cite{TorresAndres:2011wd,TorresNicola:2013}. The differences between considering the static limit ($\vec p=\vec 0$) and values of the three-momentum compatible with the temperature range considered are tiny \cite{TorresNicola:2013}, so that we will restrict ourselves to this limit to define the pion masses.

\begin{figure}
\centering
{\includegraphics[scale=0.84]{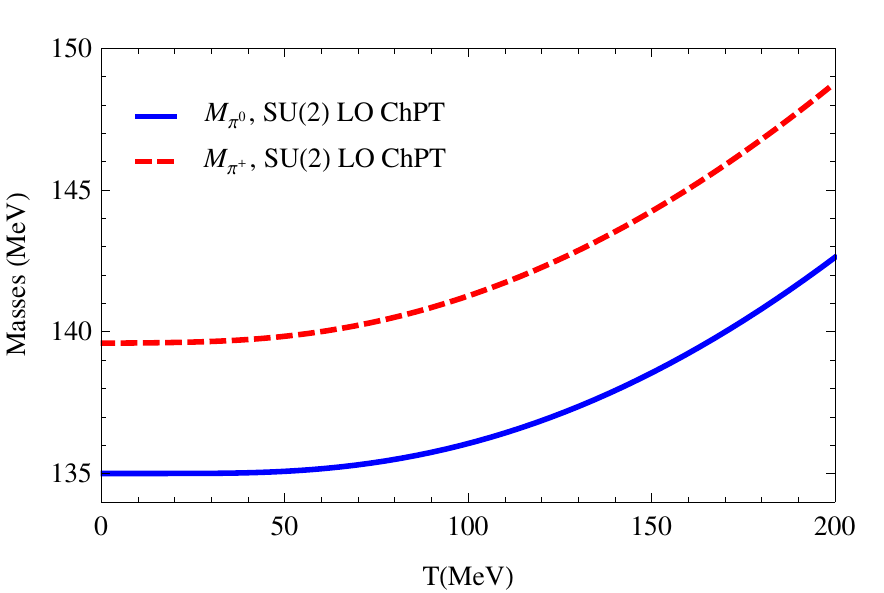}\includegraphics[scale=0.9]{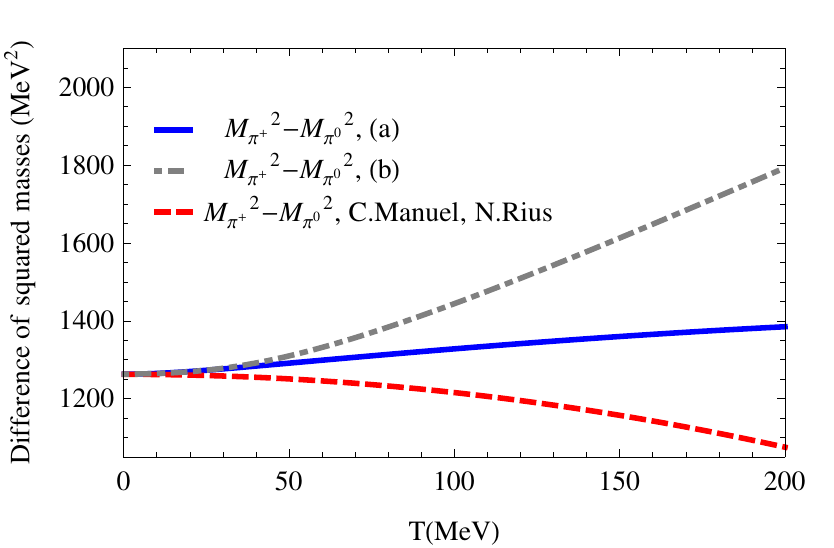}}
\caption{Left: Charged (red,dashed line) and neutral (blue,solid line) pion masses in the static limit to leading order in SU(2)-ChPT for non zero tree level pion masses. Right: Different results for the charged-neutral pion mass difference. (a, solid line) corresponds to our results in the chiral limit keeping corrections $e\neq 0$ for the tree level charged pion mass inside the loops, (b, dot-dashed line) corresponds to the same calculation with $m\neq 0$ and $e\neq 0$ also inside the loops; and the full dashed line is the result given in \cite{Manuel:1998sy}. }
\label{fig:masassu2chptstatic}
\end{figure}

Results for the charged and neutral masses separately, and for their difference can be seen in the left plot\footnote{We point out that Fig.4 and Fig.5 shown in \cite{TorresAndres:2011wd} do not correspond to the analytical expressions for the masses of the charged and the neutral pion given there, which are correct.} of Fig.\ref{fig:masassu2chptstatic}. We have used physical masses for the pions instead of the tree level masses for the numerical results, since the difference between these and the tree level ones is encoded in higher order corrections.  Despite the different sign of the various terms, the screening contribution (quadratic in $T$) turns out to dominate the pion mass difference, which, as the masses separately,  increases only slightly near the critical temperature with respect to its $T=0$ value. When $T$ grows above the applicability range of these ChPT calculations the mass difference decreases, as should be expected since expansions in $M_{\pi^\pm}/T\rightarrow 0$ should coincide with the chiral limit calculation of previous works that found a $T$-decreasing behaviour for the difference \cite{Manuel:1998sy}. For physical masses and realistic temperatures, our result is qualitatively different from the chiral limit one.

Actually, and following this last line of action, we may wonder what would our results become when taking the limit of temperatures (i) much greater than the masses and the external momenta (which means that we have to set the masses inside the loops to zero), and (ii) sizeable to the momenta, $k$, running inside the loops, i.e $T\sim k\gg m,p$; where $p$ here is the external momentum of the pion. Expanding our expressions leads to the result given in \cite{Manuel:1998sy},
$M^2_{\pi^\pm}-M^2_{\pi^0}=\hat M^2_{\pi^\pm}\left(1-\frac{T^2}{6}\right)+\frac{1}{4}e^2T^2,$
which serves us as another consistency check.

As commented above, our low temperature analysis allows to work with a slightly different chiral limit, in the sense that we can still assume $m=0$ but considering $e\neq 0$, even inside the loops. In the right plot of Fig.\ref{fig:masassu2chptstatic} we show our calculation both in this latter limit, and also considering $m\neq 0$, $e\neq 0$; to be compared with those appearing in the chiral limit result \cite{Manuel:1998sy}. The screening-like terms, always increasing with $T$ and inherent to the thermal bath, compensate for the  terms coming from the sum rule relating the axial and vector spectral functions \cite{Das:1967it}, which should decrease applying chiral symmetry restoration arguments \cite{Dey:1990ba}. The sum rule is not valid for non-zero quark masses, but one can still identify the contributions from the vector and axial spectral functions from a resonance exchange model, as we will briefly describe in the next section.

\section{Resonance contributions}

The results for the pion mass difference in ChPT is quantitatively reliable only in a temperature range up to, at most, 100-150 MeV. One way of increasing the predictive quality of the calculations and check the validity range of the ChPT result -at the risk of losing the model-independent character of the theory- is to use a model \cite{ Ecker:1988te,Ecker:1989yg} that implements the couplings with the vector resonance  $J^{PC}=1^{--}$  and the axial  $J^{PC}=1^{++}$. This allows to calculate the corrections to the charged pion mass (neutral pions do not couple to this interactions in the model) allowing to extend the calculations up to intermediate energies ($\sim$1 GeV) \cite{Donoghue:1996zn}.

\begin{figure}
\centering
\includegraphics[scale=0.14]{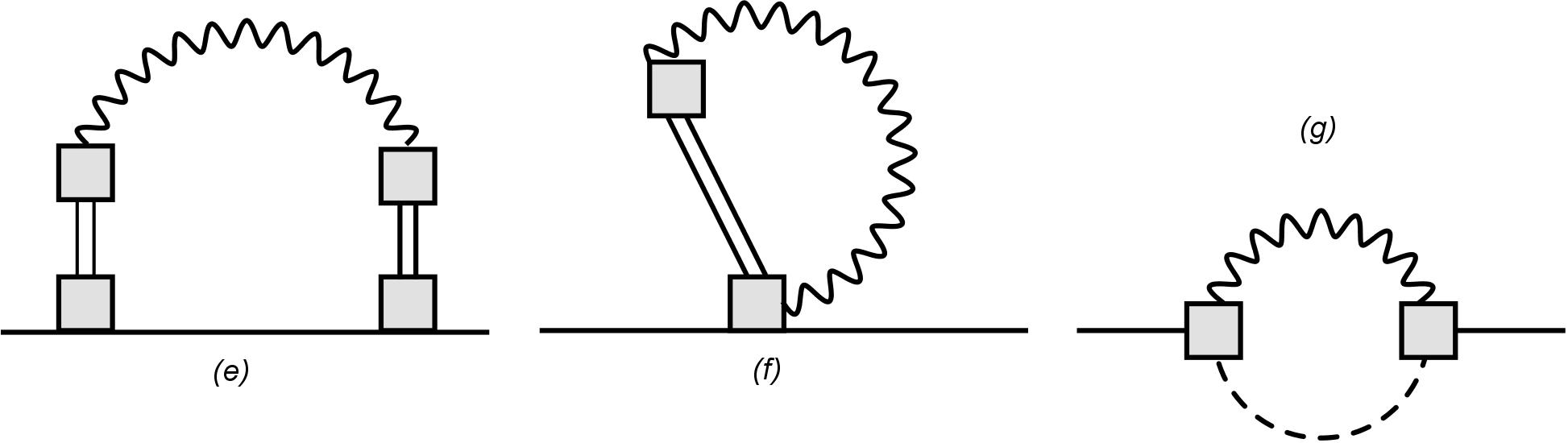}
\caption{1-PI diagrams contributing at leading order to the charged pion-self energy. $\rho$ and a$_1$ particles are represented by double and dashed lines, respectively. The relevant vertices including charged pions, resonances and photons are drawn as grey boxes.}
\label{fig:diagres}
\end{figure}

The couplings with the spin-1 resonance fields relevant for the leading order calculation of the charged pion self-energy are diagrammatically expressed in Fig.\ref{fig:diagres}. It is convenient to note that we still have the contributions coming from charged and neutral pion tadpoles and virtual photon diagrams. Since resonance fields only couple to charged pions, the neutral pion mass at leading order in this resonance model is the same as that obtained in the previous section.

As commented before, at $T=0$ and in the chiral and soft pion limit, it was shown \cite{Das:1967it} that the pion mass difference can be parametrized exactly in terms of the difference of the vector-axial spectral functions, so the corrections to this formula must be proportional to $\hat M_\pi$, as was  shown in \cite{Baur:1995ig} using ChPT at zero temperature. The modifications to the expression of Das et al. for non-zero quark masses and temperatures are discussed in \cite{TorresNicola:2013}.

Although there are more realistic calculations making a non-narrow width treatment of the resonances \cite{Donoghue:1996zn}, we will suppose that the spectral functions $\rho_V(s)$ and $\rho_A(s)$ are of the form $\rho_{V,A}(s)=F^2_{V,A}\delta(s-M^2_{V,A})$, where $F_V$, $F_A$, $M_V$ and $M_A$ stand for the vector (axial) decay constants; and vector (axial) masses, respectively. We will use the numerical values for this constants given in \cite{Donoghue:1996zn}. The thermal modification of these spectral functions amounts to leading order to the pion tadpole contributions and accounts for the $\rho-a_1$ mixing in the thermal bath \cite{Dey:1990ba}. Other corrections -such as the modification of the $\rho$ width showing up in the dilepton spectrum- are  suppressed in the pion mass difference for the temperature range considered here.

\begin{figure}
\centering
{\includegraphics[scale=0.81]{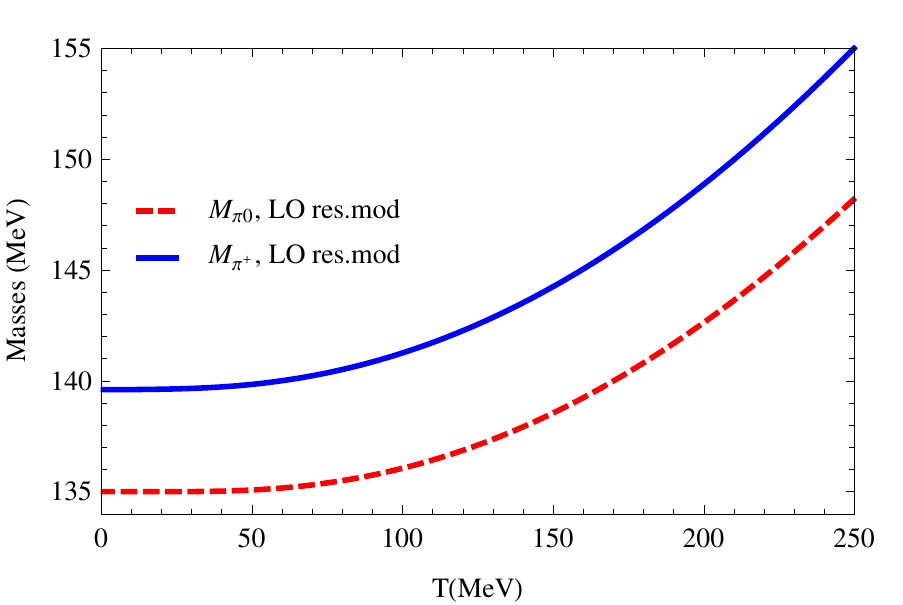}\includegraphics[scale=0.88]{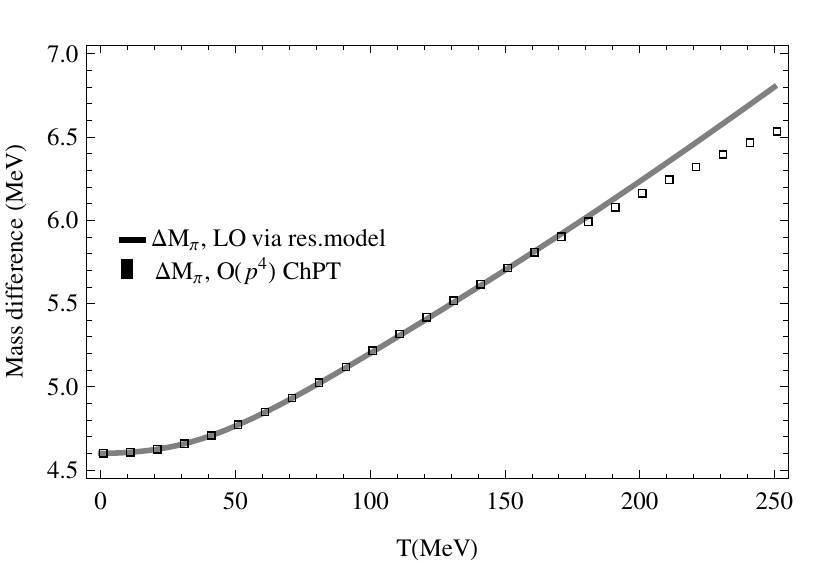}}
\caption{Charged and neutral pion masses in the static limit to leading order in the light resonance model (left plot), and their difference (right plot, solid line) compared to the leading order ChPT calculation (right plot, squares).}
\label{fig:contmpichres}
\end{figure}

The results for the charged pion and the neutral masses in the static limit, as well as the pion mass difference in this model compared with that calculated in SU(2)-ChPT to leading order, are shown in Fig.\ref{fig:contmpichres}. The first we observe is that contributions coming from the interaction with resonances are only activated at temperatures much higher than those at where low energy effects start to be important (which correspond to the ChPT results), as we would have expected since, up to $T\sim 200$ MeV, resonances are suppressed. The sub-leading resonance corrections are typically of $\Od (T^2 M_\pi^2/M_{\rho,a_1}^2)$.  Due to this very same fact, the contribution corresponding to the exchange of an a$_1$ and a photon is much smaller than those in which the $\rho$ interacts (form factor and tadpole-like diagrams). It turns out that the inclusion of resonant contributions does not change significantly the ChPT result (extrapolating the thermal range in which results are realistic, we get approximately $5\%$ of discrepancy at 250 MeV). Of course, the lack of an appropriate coupling constant for perturbative calculations implies the fact that there may exist other higher order diagrams which can be numerically relevant at finite temperature, and we must not exclude these resonances to have a more important role even in this intermediate energy region, as they have at zero temperature \cite{Das:1967it}.

\end{document}